\documentstyle[preprint]{ptptex}
\input epsf.sty
\preprintnumber[3cm]{
KOBE-TH-99-04\\
hep-th/9905100}
\renewcommand{\L}{{\cal L}}
\newcommand{\mass}{M_{\mathrm{H}}}
\newcommand{\Veff}{V_{\mathrm{eff}}}
\newcommand{\Nf}{N_{\mathrm{f}}}
\newcommand{\Nad}{N_{\mathrm{ad}}}
\newcommand{\Nr}{N_{\mathrm{R}}}
\renewcommand{\Re}{\mbox{Re}}
\newcommand{\Li}{\mbox{Li}}
\newcommand{\diag}{\mbox{diag}}
\title{
Matter Representations and \\
Gauge Symmetry Breaking via Compactified Space}

\author{
 Hisaki {\sc Hatanaka}
 \footnote{E-mail: hatanaka@phys.sci.kobe-u.ac.jp}
}

\inst{
Graduate School of Science and Technology, Kobe University \\
Kobe 657-8501, Japan
}

\recdate{March 5, 1999}

\abst{
We study dynamical gauge symmetry breaking via compactified space in the
framework of $SU(N)$ gauge theory in $M^{d-1}\times S^1$ ($d=4,5,6$)
space-time. In particular, we study in detail the gauge symmetry
breaking in $SU(2)$ and $SU(3)$ gauge theories when the models contain
both fundamental and adjoint matter.  As a result, we find that any
pattern of gauge symmetry breaking can be realized by selecting an appropriate
set of numbers $(\Nf,\Nad)$ in these cases. This is achieved without
tuning boundary conditions of the matter fields. As a by-product, in some
cases we obtain an effective potential which has no curvature at the minimum,
thus leading to massless Higgs scalars, irrespective of the size of
the compactified space.}

\begin{document}

\maketitle

%%%%%%%%%%%%%%%% section 1 %%%%%%%%%%%%%%%%%%

\section{Introduction}

\subsection{
The gauge hierarchy problem and higher-dimensional gauge theories }

We have proposed a possibility of solving the gauge hierarchy
problem\cite{HIL98} by invoking higher-dimensional gauge symmetry.

Our main idea in this scenario is the following. Suppose we consider a
gauge theory on the $d$-dimensional manifold $M^{\mathrm{Minkow}} \times
M^{\mathrm{extra}}$, where $M^{\mathrm{Minkow}}$ is an ordinary
Minkowskian 4-dimensional space-time, and $M^{\mathrm{extra}}$ is a
compactified spatial $(d-4)$-dimensional manifold. The gauge fields
$A_M$ on $M^{\mathrm{Minkow}} \times M^{\mathrm{extra}}$ have $4+(d-4)$
components:
\begin{equation} 
A_{\mu} (\mu=0, \cdots ,3), \,\,\,   A_m(m= 4, \cdots ,d-1).
\end{equation}
If we regard zero-modes of $d-4$ components of gauge fields as ``Higgs
scalars'', we can avoid quadratically divergent quantum corrections of
Higgs boson masses thanks to the higher-dimensional gauge symmetry.

In a previous work,\cite{HIL98} we considered a QED model on
$M^{d-1}\times S^1$ space-time and took a zero-mode of $d$-th the
component of the gauge field as a ``Higgs scalar''. We calculated also
the quantum mass correction of the $d$-th component of the gauge
field. As a result we confirmed the disappearance of the quadratic
divergence.  We, however, have found that a finite mass correction to
the ``Higgs'' remains, which is generally proportional to the inverse of
the $S^1$ radius.

Although the gauge hierarchy problem in its original sense has been
solved, as the quadratic divergence has disappeared, several issues have
to be settled for the scenario to be realistic. Here we address the
following two important issues. (i) In our toy model on $M^4 \times
S^1$, the Higgs mass $\mass$ experiences a finite but generally large
quantum correction provided that the radius of $S^1$ is as small as the
GUT or Planck length.\cite{HIL98} Then the issue is how we can make the
finite correction to $\mass$ small, approximately comparable to the weak
scale.  (ii) The toy model on $M^4 \times S^1$ is higher dimensional
QED.  The $U(1)$ gauge symmetry, therefore, is not spontaneously broken,
even though the Higgs $d$-th component of the photon field has
non-vanishing vacuum expectation value (VEV). This is simply because the
Higgs field is electrically neutral, and there appears the possibility
to break the gauge symmetry, once the theory is extended to the
non-Abelian gauge (Yang-Mills) theory, where the Higgs behaves as an
adjoint representation of gauge group.

With regard to the first issue above, we have already discussed a few
possibilities to settle the problem in a previous work.\cite{HIL98} The
first rather trivial possibility is to assume a ``large''
compactification scale such as TeV$^{-1}$. \cite{Antoniadis}\cite{Large} 
It is interesting to note that such a large scale compactification has
attracted revived interest, partially inspired by the progress in
superstring/M theories.\cite{Horava96} The second possibility is to
utilize a heavy electron where mass $m$ satisfies $mL \gg 1$ ($L$:
circumference of $S^1$). Then the correction to $\mass$ is suppressed
roughly as $\exp(-mL)/L$. (Here $\exp(-mL)$ corresponds to the Boltzmann
factor in finite temperature field theory.) It has also been pointed out
that the finite mass $\mass$ can be regarded as coming from a sort of
Aharonov-Bohm effect on $S^1$ space and may disappear if the
compactified space is a simply-connected manifold. We have confirmed
this expectation in a toy model on $S^2$, showing that $\mass$
identically vanishes.

The main purpose of this paper is to settle the remaining second issue
mentioned above. In theories with $S^1$ as the extra space, the constant
background and VEV of gauge fields have physical meaning as a Wilson
loop, and they can be dynamically fixed by the minimization of effective
potential as functions of the background fields. Thus in Yang-Mills
theories the VEV behave just as the VEV of adjoint Higgs, and they
generally break the gauge symmetry spontaneously: the so-called
``Hosotani mechanism''.\cite{Hosotani83}

To obtain realistic GUT-type theories, starting from higher-dimensional
Yang-Mills theories, it is quite important to study how the gauge
symmetry is dynamically broken by the Hosotani mechanism. In particular,
we are interested in the question of how the breaking pattern of gauge
symmetry depends on the choice of the representation (fundamental or
adjoint or $\cdots$) of matter fields. We will extensively study this
point by considering two representative gauge groups, $SU(2)$ and
$SU(3)$. As a bonus we also find an interesting property of the theory,
concerning the first issue mentioned above. Namely, we realize in some
cases that the curvature of the curve of the potential function
identically vanishes, leading to $\mass=0$. It should be noted that this
vanishing $\mass$ is obtained for arbitrary compactification scale and
without the introduction of a simply-connected space like $S^2$.
\cite{HIL98}

\subsection{
Gauge symmetry breaking and matter representation in Hosotani mechanism}

We now briefly review the history of the investigation of the gauge
symmetry breaking via the Hosotani mechanism, and in particular the case
of $M^{d-1} \times S^1$ space-time configuration.

In 1983, Hosotani found that a gauge symmetry may break down in a system
of $SU(N)$ gauge theories on $M^3 \times S^1$ space-time.  But in his
model in order to break the gauge symmetry, it was necessary to have
both many (more than 16 flavors) complex scalar fields as the matter
field and a non-trivial boundary conditions for the matter fields,
$\delta=\pi/2$.\cite{Hosotani83} Later it became obvious that for the
case in which fermions belong to the fundamental representation of the
$SU(N)$ gauge group, gauge symmetry is never broken. In 1989, It was
found that $SU(2)$ gauge symmetry breaking is possible when fermions are
in an adjoint representation,\cite{Higuchi88}\cite{Davies88} and it was
shown that the pattern of gauge symmetry breaking depends on the {\it
congruency class representation of the gauge group}.\cite{Davies88}
\cite{Davies89}

Many works have been done along this direction, giving hope for using
the Hosotani mechanism as a possible mechanism of GUT symmetry
breaking. This scenario has been investigated to some extent.
\cite{Hosotani89}\cite{hohosotani}
However, the case in which the theory contains both fundamental and
adjoint representations of fermions has never been investigated, except
the $E_6$ model in Ref. \cite{Davies88} and supersymmetric models.
\cite{Murayama}\cite{Takenaga} I was partially motivated by these works
in supersymmetric models and attempt to study the non-supersymmetric
version of gauge symmetry breaking via the Hosotani mechanism in theories
with both fundamental and adjoint matter.

\section{Dynamical gauge symmetry breaking through
extra dimension}

\subsection{Effective Potential}

We consider $SU(N)$ gauge theories in $M^{d-1} \times S^1$ space-time
with both fundamental and adjoint fermionic fields.

The Lagrangian is given by
\begin{eqnarray}
\L
 &=& -\frac{1}{2} \mbox{Tr} F^{MN}F_{MN}
     + \sum_{j=1}^{\Nf}  \bar{\psi}_{{\mathrm{f}},j } 
       i\gamma^M D_M \psi_{{\mathrm{f}},j} 
     + \sum_{k=1}^{\Nad} \bar{\psi}_{{\mathrm{ad}},k} 
       i\gamma^M D_M \psi_{{\mathrm{ad}},k} , 
\end{eqnarray}
where
\begin{eqnarray}
 F_{MN}
 &=& \partial_M A_N - \partial_N A_M + ig [ A_M , A_N ],
\,\,\,\,\,\,\,\,\,\, A_M = A_M^a T^a ,
\\
D_M \psi_{\mathrm{f}}
 &=& \partial_M \psi_{\mathrm{f}} + ig T^a A_M^a \psi_{\mathrm{f}}, \\
D_M \psi_{\mathrm{ad}}
 &=& \partial_M \psi_{\mathrm{ad}} + ig [T^a A_M^a, \psi_{\mathrm{ad}}] ,
\end{eqnarray}
$\psi_{\mathrm{f}}$($ \psi_{\mathrm{ad}}$) are fermions in the
fundamental (adjoint) representation, and $\Nf$ ($\Nad$) are the flavor
numbers of the Dirac fermion in the fundamental (adjoint)
representation. The indices $M$ and $N$ run from $0$ to $d-1$ (whereas
$\mu$ and $\nu$ run from $0$ to $d-2$) and $j$ and $k$ are flavor
indices. The color index has been omitted here. We use $x$ and $y$ as
the coordinates on $M^{d-1}$ and $S^1$, respectively.

The boundary conditions on $S^1$ are taken as
\begin{eqnarray}
A_\mu(x,y+L)           &=& A_\mu(x,y), \\
\psi_{\mathrm{f}}(x,y+L)    &=& e^{i\delta_{\mathrm{f}}}  
                                \psi_{\mathrm{f}}(x,y), \\
\psi_{\mathrm{ad}}(x,y+L)   &=& e^{i\delta_{\mathrm{ad}}}
                                \psi_{\mathrm{ad}}(x,y),
\end{eqnarray}
where $L$ is the compactification scale of the $S^1$ subspace and is
related to the radius of $S^1$ as $L=2\pi R$.

The effective potential is separated into three parts:
\cite{Hosotani89}\cite{hohosotani}
\begin{eqnarray}
\Veff
&=&
\Veff^{\mathrm{g+gh}}
+\Veff^{\mathrm{f}}
+\Veff^{\mathrm{ad}},
\\
\Veff^{\mathrm{g+gh}}
&=&
- (d-2)\frac{\Gamma(d/2)}{\pi^{d/2}L^d}
\sum_{i,j=1}^N \sum_{n=1}^\infty \frac{\cos[n(\theta_i - \theta_j)]}{n^d},
\\
\Veff^{\mathrm{f}}
&=&
\Nf \cdot 2^{[d/2]}\frac{\Gamma(d/2)}{\pi^{d/2}L^d}
 \sum_{i=1}^N \sum_{n=1}^\infty \frac{\cos[n(\theta_i-\delta_{\mathrm{f}})]}{n^d}
\label{eq:Vf},
\\
\Veff^{\mathrm{ad}}
&=&
\Nad \cdot 2^{[d/2]}\frac{\Gamma(d/2)}{\pi^{d/2}L^d}
 \sum_{i,j=1}^N \sum_{n=1}^\infty 
 \frac{\cos[n(\theta_i - \theta_j - \delta_{\mathrm{ad}})]}{n^d}
\label{eq:Vad}.
\end{eqnarray}
Here $\Veff^{\mathrm{g+gh}}$, $\Veff^{\mathrm{f}}$ and
$\Veff^{\mathrm{ad}}$ are the contributions from the loops of gauge and
ghost fields, fundamental fermions and adjoint fermions, respectively.
The $\theta_i$ ($i= 1 \cdots N$) denote the components of the diagonalized
constant background gauge field (or VEV of the gauge field):
\begin{eqnarray}
\langle A_y \rangle = \frac{1}{gL}
\pmatrix{
\theta_1 & & \cr
         & \ddots & \cr
         &        & \theta_N},
\,\,\,\,\,\,\,\,
\sum_{i=1}^N \theta_i=0 .
\end{eqnarray}
The $\theta_i$ are called {\it the non-integrable phases of the gauge
field}\cite{Hosotani89} and cannot be constrained by the pure-gauge
condition $F_{MN}\equiv 0$. They should be determined dynamically at the
quantum level as a result of the minimization of $\Veff$.

When non-zero $\theta_i$ are induced dynamically, only generators of
the gauge group, which commute with the Wilson line surrounding $S^1$,
\begin{eqnarray}
\langle W(x) \rangle 
\equiv {\cal P} \exp 
\left[ig 
\int_{x,y}^{x,y+L} \langle A_y \rangle dy \right]
= \pmatrix{\exp i\theta_1 & & \cr & \ddots & \cr & & \exp i\theta_N},
\end{eqnarray}
form a remaining subgroup of $SU(N)$, and other symmetries are
broken. Thus gauge symmetry breaking is induced through dynamics in the
extra dimension.

The phases $\delta^{\mathrm{f}}$ and $\delta^{\mathrm{ad}}$, relevant
for the fermions' boundary conditions, remain as non-dynamical
parameters of the model and must be set by hand.  In the $\Nad \neq 0$
and $\Nf=0$ cases, \cite{Hosotani89} \cite{hohosotani} the gauge
symmetry breaking is controlled by changing the boundary condition
$\delta_{\mathrm{ad}}$s. In this paper, however, we show that we can
also realize various patterns of gauge symmetry breaking by varying the
appropriate choice of $\Nf$ and $\Nad$, even if we simply take the
periodic boundary condition of fermions,
\begin{eqnarray}
\delta_{\mathrm{f}} = \delta_{\mathrm{ad}} = 0.
\end{eqnarray}

When the periodic boundary condition is taken for fermions, $\Veff$ is
simplified as
\begin{eqnarray}
\Veff = \frac{\Gamma(d/2)}{\pi^{d/2}L^d}
\left\{
\left[-(d-2) + 2^{[d/2]} \Nad \right]
  \sum_{i,j=1}^N F_d(\theta_i-\theta_j)
+2^{[d/2]} \Nf \sum_{i=1}^N F_d(\theta_i)
\right\},
\end{eqnarray}
where the function $F_d(\theta)$ is defined as
\begin{eqnarray}
F_d(\theta)
 \equiv \sum_{n=1}^\infty \frac{\cos n\theta}{n^d} .
\end{eqnarray}
It can also be written in terms of the {\it polylogarithmic functions}
$\Li_d(x)\equiv \sum_{n=1}^{\infty}\frac{x^n}{n^d}$:
\begin{eqnarray}
F_d(\theta)
= \Re \sum_{n=1}^\infty \frac{\exp(in\theta)}{n^d}
= \Re\, \Li_d(\exp(i\theta)).
\end{eqnarray}
Thus we can calculate $F_d(\theta)$ analytically for some special values,
or even numerically with arbitrary precision.

It is easy to show how each sector of the effective potential plays a
role in the gauge symmetry breaking.  The combination of gauge and
fundamental fermion sectors has global minima in a symmetric
configuration of $\theta$, where $\langle W \rangle$ is $SU(N)$ gauge
symmetric.  On the other hand, the adjoint fermion sector has global
minima at another $\theta$, where $\langle W \rangle$ allows only
$U(1)^N$ symmetry.  Thus we can expect that by changing $\Nf$ and $\Nad$
we get a variety of gauge symmetry breaking.

Some comments are now in order.

%\begin{description}
%\item[Effects of scalar field]
\subsubsection{Effects of scalar field}

It is obvious that if there are also some scalars in the theory, the
cotribution of the $N_{\mathrm{f,ad}}^{\mathrm{scalar}}$ flavor complex
scalar fields to the effective potential is given by replacing
$2^{[d/2]}N_{\mathrm{f,ad}}$ with $ - 2
N_{\mathrm{f,ad}}^{\mathrm{scalar}}$. Thus, $\Nf$ and $\Nad$ in
Eqs.~(\ref{eq:Vf}) and (\ref{eq:Vad}) should be understood as
\begin{eqnarray}
\Nr  = \Nr^{\mathrm{fermion}} - \frac{2}{2^{[d/2]}} \Nr^{\mathrm{scalar}},
\end{eqnarray}
where R stands for the representation and $\Nr^{\mathrm{fermion}}$ and
$\Nr^{\mathrm{scalar}}$ are the number of flavors of fermions and
scalars in the representation R, respectively. Therefore $\Nf$ and
$\Nad$ can be negative and even take rational number values.

Our investigation can be applied to any region of $(\Nf,\Nad)$ plane.
The $\Nad < (d-2)/2^{[d/2]}$ region is investigated without loss of
generality by investigating the $\Nad=0$ case and has been already
investigated by many authors. The region defined by $\Nad >
(d-2)/2^{[d/2]}$ and $\Nf=0$ has also been investigated,\cite{Higuchi88}
\cite{Davies88} \cite{Davies89} \cite{Hosotani89} \cite{hohosotani} and
it has been claimed that the gauge symmetry breaking is mainly handled
by the boundary conditions of fermion fields.\cite{Hosotani89}
\cite{hohosotani} But the region defined by $\Nad > (d-2)/2^{[d/2]}$ and
$\Nf \neq 0$ has not yet been investigated. Hence we study what happens
if both adjoint and fundamental matter are included in the Hosotani
mechanism.

%\item[Supersymmetry]
\subsubsection{Supersymmetry}

If Majorana, Weyl or Majorana-Weyl fermions exist in $d$-dimensional
space-time and if we take such a type of fermion as the matter field,
instead of the Dirac fermion, $\Nr$ should be replaced by $\frac{1}{2}
\Nr^{\mathrm{Weyl}}$, $ \frac{1}{2}\Nr^{\mathrm{Majorana}}$ or
$\frac{1}{4} \Nr^{\mathrm{Majorana\textrm{-}Weyl}}$, respectively. It
also must be pointed out that if the specific relations $2^{[d/2]}\Nad -
(d-2) = 0$ and $\Nf=0$ hold, and all $\Nad$ matter fields are fermionic,
the fermionic fields can be regarded as ``gaugino'', and the system may
be regarded as a super Yang-Mills theory. For supersymmetric pure
Yang-Mills theories, however, the effective potentials calculated in
this article all vanish and will not be investigated further. When the
boundary conditions of bosons and fermions are different, this is not
the case, and symmetry breaking results even if $\Nad = (d-2) /
2^{[d/2]}$ and $\Nf=0$. This is related to the Scherk-Schwartz mechanism
of supersymmetry breaking, \cite{SS}\cite{Murayama}\cite{Takenaga} and
we will present the results of an investigation along this line in the
near future.
%\end{description}

\subsection{$SU(2)$ case}
In the $SU(2)$ case, the parameterization of background gauge fields is
given as
\begin{eqnarray}
\langle A_y \rangle gL
=\diag(\theta , -\theta).
\end{eqnarray}
The $SU(2)$ symmetric gauge configuration corresponds to $\langle A_y
\rangle gL = \diag(\pi,-\pi)$ or $\diag(0,0) $.  Hence, when
$\Veff(\theta)$ has a minimum at $\theta=\pi$ or $\theta=0$, the $SU(2)$
symmetry remains dynamically.

We search for global minima of $V(\theta)$ for various $\Nf$ and $\Nad$
with periodic boundary conditions
($\delta_{\mathrm{f}}=\delta_{\mathrm{ad}} =0$). To see the contribution
of each sector, $\Veff^{\mathrm{g+gh}}, \Veff^{\mathrm{f}}$ and
$\Veff^{\mathrm{ad}}$ are plotted in Figs.1(a),(b) and (c), respectively.
\begin{figure}[htbp]
%\begin{center}
%(a) \resizebox{6cm}{!}{\includegraphics{su2g.eps}} \\ %\hspace{1cm}
%(b) \resizebox{6cm}{!}{\includegraphics{su2f.eps}} \\ %\hspace{1cm}
%(c) \resizebox{6cm}{!}{\includegraphics{su2ad.eps}}
%\end{center}
\epsfxsize=5cm
\centerline{(a)\epsfbox{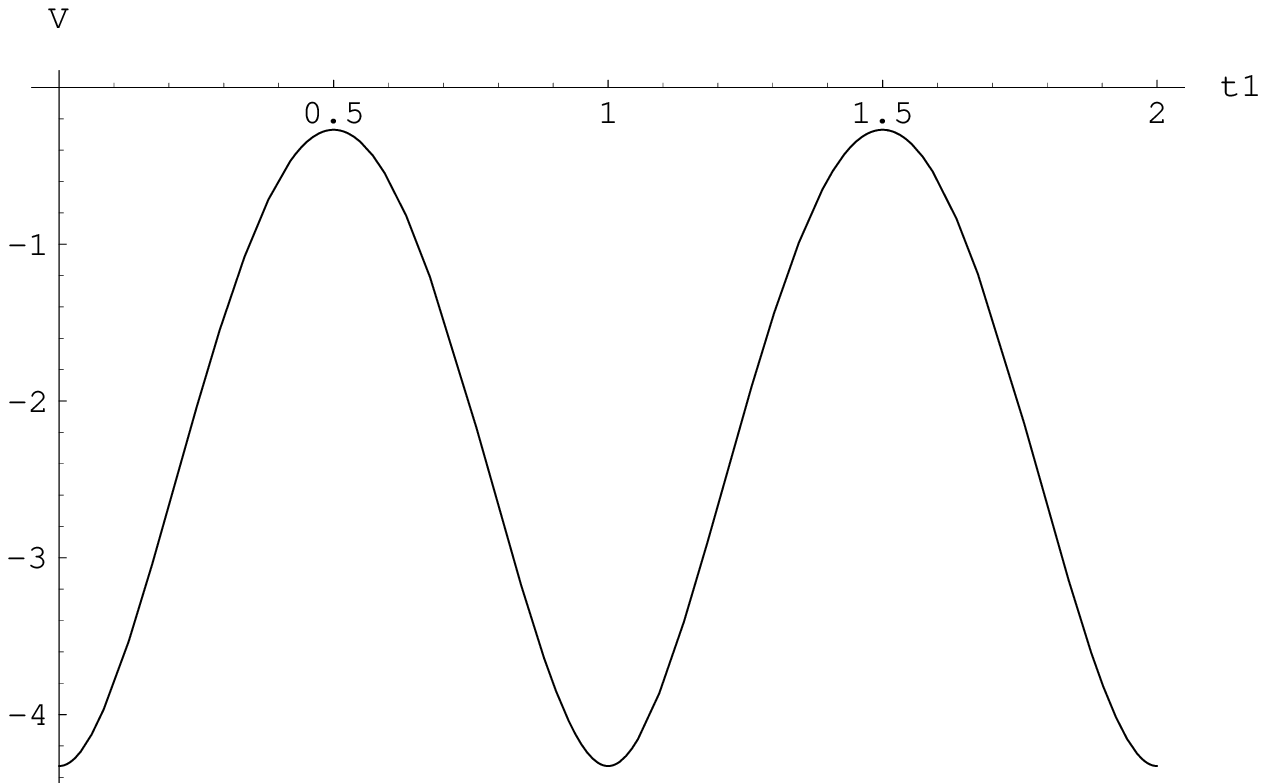}}
\epsfxsize=5cm
\centerline{(b)\epsfbox{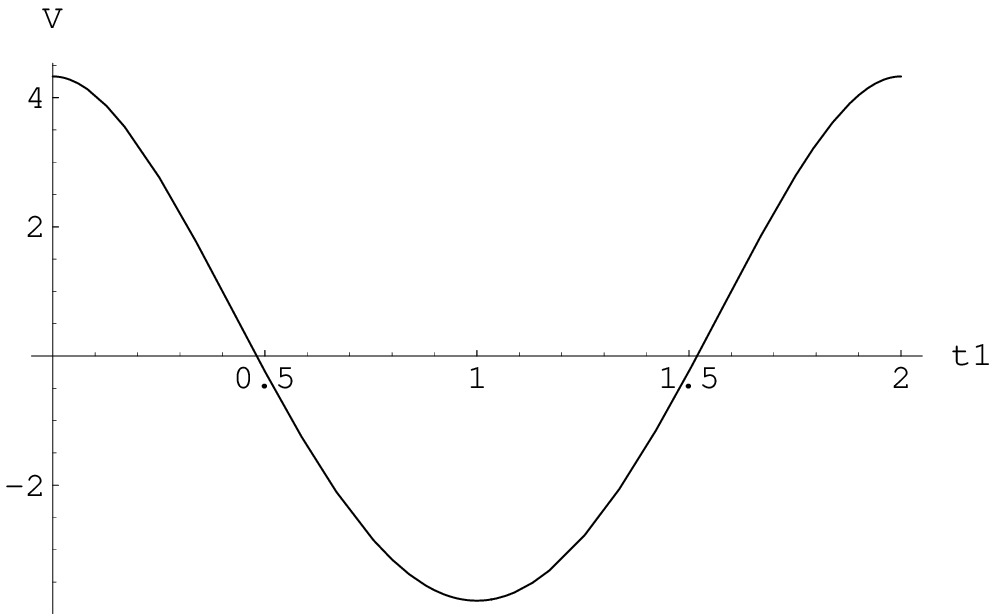}}
\epsfxsize=5cm
\centerline{(c)\epsfbox{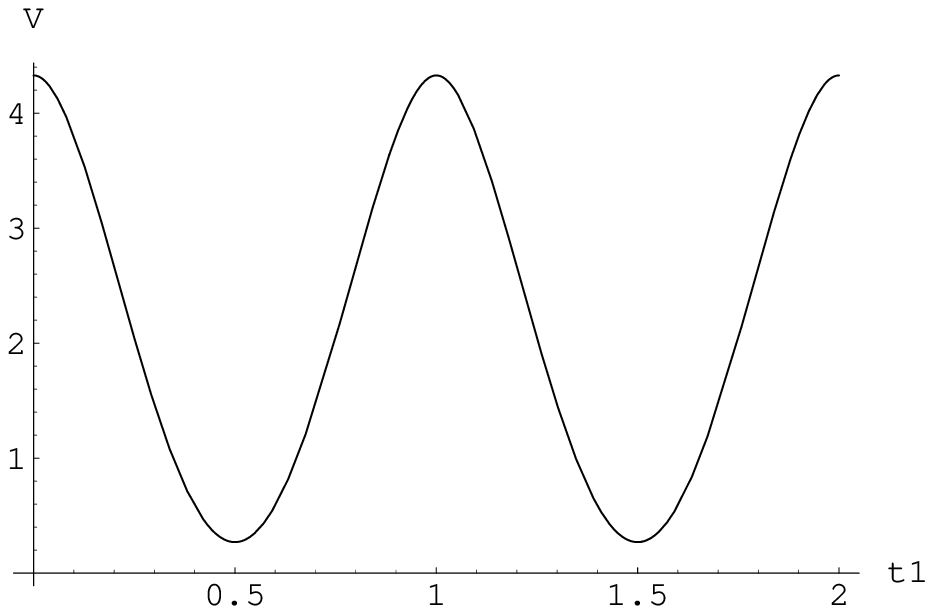}}
\caption{Plots of $\Veff$ as a function of $\theta_1$ (indicated as t1).
 (a) $\Veff^{\mathrm{g+gh}}$ (for $\Nad=\Nf=0$),
 (b) $\Veff^{\mathrm{f}}$    (for $\Nad=1/2$, $\Nf=1$), and
 (c) $\Veff^{\mathrm{ad}}$   (for $\Nad=1$, $\Nf=0$).}
\end{figure}
In these figures, we find that the points $\theta=0$ and $\pi$ always
become global minima of $\Veff^{\mathrm{g+gh}} + \Veff^{\mathrm{f}}$. On
the other hand, $\Veff^{\mathrm{ad}}$ has global minima at $\theta =
\pi/2$ and $3\pi/2$, at which the $SU(2)$ symmetry is broken. Thus to
judge whether there appears another local minimum to break $SU(2)$ that
is lower than those at $\theta=0$ and $\pi$, it is useful to calculate
the second derivative of $\Veff$ at $\theta=0$ and $\pi$. If $SU(2)$
starts to break at some number of $\Nf$ and $\Nad$, the second
derivative at $\theta=0$ or $\pi$ is expected to vanish for such a pair
of numbers. For the specific case of $d=4$, the calculated second
derivatives are found to behave as
\begin{eqnarray}
\left.
\frac{\partial^2 \Veff^{d=4}}{\partial \theta^2}
\right|_{\theta=\pi}
&\propto&
\Nf - 8\Nad +4,
\\
\left.
\frac{\partial^2 \Veff^{d=4}}{\partial \theta^2}
\right|_{\theta=0}
&\propto&
2 - \Nf -4 \Nad,
\end{eqnarray}
where we have used $F''_4(0) = -\pi^2 / 6$ and $F''_4(\pi) =
\pi^2/12$. Thus the conditions $\Nf - 8\Nad + 4 = 0$ and $2 - \Nf -
4\Nad = 0$ give critical numbers for $\Nf$ and $\Nad$. In this way, the
breaking pattern of $SU(2)$ for given $\Nf$ and $\Nad$ is known to be
\begin{equation}
SU(2) \rightarrow \left\{
\begin{array}{ccccccc}
SU(2)  & \mbox{for} &            &    & \Nf &\leq& 2 - 4\Nad,  \\
U(1)   & \mbox{for} & 2 - 4 \Nad &<   & \Nf &<   & 8\Nad-4, \\
SU(2)  & \mbox{for} & 8\Nad-4    &\leq& \Nf. &    &
\end{array}
\right.\label{eq:su2pat}
\end{equation}

It is quite interesting to note the fact that when equality is satisfied
in (\ref{eq:su2pat}), not only is $SU(2)$ symmetry restored, but also the
second derivative of gauge potential is exactly 0, which means that the
Higgs scalar, as an extra gauge boson, has a vanishing mass for arbitrary
$L$, even in the model on $S^1$.

\subsubsection{$d=5$, $d=6$ case}

Using $ F''_5(0) =F''_5(2\pi n) = -\zeta(3)$, $F''_5(\pi)=3\zeta(3)/4$,
$F''_6(0) =F''_6(2\pi n) = -\zeta(4) = - \pi^4 / 90$,$F''_6(\pi) =
7\pi^4 / 720$, where $\zeta(\alpha)$ is Rieman's zeta function, we can
also determine the symmetry breaking pattern for the $d=5$ and $d=6$ cases.
For $d=5$,
\begin{equation}
\mbox{$SU(2)$}
\rightarrow 
\left\{
\begin{array}{ccccccc}
SU(2) & \mbox{for} &               &     & \Nf &\leq& 3-4 \Nad, \\
U(1)  & \mbox{for} & 3-4\Nad       &<    & \Nf &<   & (16\Nad-12)/3, \\
SU(2) & \mbox{for} & (16\Nad-12)/3 &\leq & \Nf, &    &
\end{array}
\right. 
\end{equation}
and for $d=6$,
\begin{equation}
SU(2)
\rightarrow 
\left\{
\begin{array}{ccccccc}
SU(2) & \mbox{for} &               &    & \Nf &\leq& 2 - 4 \Nad, \\
U(1)  & \mbox{for} & 2 - 4\Nad     &<   & \Nf &<   & (32\Nad - 16)/7, \\
SU(2) & \mbox{for} & (32\Nad-16)/7 &\leq& \Nf. &    &
\end{array}
\right. 
\end{equation}

We thus have found that in both cases, gauge symmetry breaking and its
restoration occur, though the critical values of $(\Nf,\Nad)$ are
different.

\subsection{$SU(3)$ case}

We take the parameterization of the background $SU(3)$ gauge field as
\footnote{We can also use another parameterization such as $\langle A
\rangle gL = \theta_1 \lambda_3 + \theta_2 \lambda_8$, where the $\lambda_i$
are Gell-Mann's $\lambda$ matrices. But the following discussion does
not depend on the choice of parameterization.}
\begin{eqnarray}
\langle A_y \rangle gL =
\diag(\theta_1 , \theta_2 , -(\theta_1+\theta_2)).
\end{eqnarray} 
The case in which $SU(3)$ is exact and not broken is realized for
\begin{eqnarray}
\langle A_y \rangle gL &=& \diag(0 , 0 , 0), \\
\langle A_y \rangle gL &=& \diag(2\pi/3 , 2\pi/3 , -4\pi/3)
\,\,\,
\mbox{and its permutations.}
\end{eqnarray}
The symmetry breaking of $SU(3)$ into $SU(2)\times U(1)$ is realized
provided
\begin{eqnarray}
\langle A_y \rangle gL = \diag(0 , \pi , -\pi)
\,\,\,
\mbox{and its permutations,}
\end{eqnarray}
or generally,
\begin{eqnarray}
\langle A_y \rangle gL = \diag(\theta , \theta , -2\theta)
\,\,
(\theta \neq -2\theta \bmod 2\pi) \,\, \mbox{and its permutations.}
\end{eqnarray}

The contribution of each sector to $\Veff$ is shown in Fig. 2.
\begin{figure}[htbp]
%\begin{center}
%(a) \resizebox{6cm}{!}{\includegraphics{su3g.eps}} \\  %\hspace{0.5cm}
%(b) \resizebox{6cm}{!}{\includegraphics{su3f.eps}} \\  %\hspace{0.5cm}
%(c) \resizebox{6cm}{!}{\includegraphics{su3ad.eps}}
%\end{center}
%(a)
\epsfxsize=5cm
\centerline{(a)\epsfbox{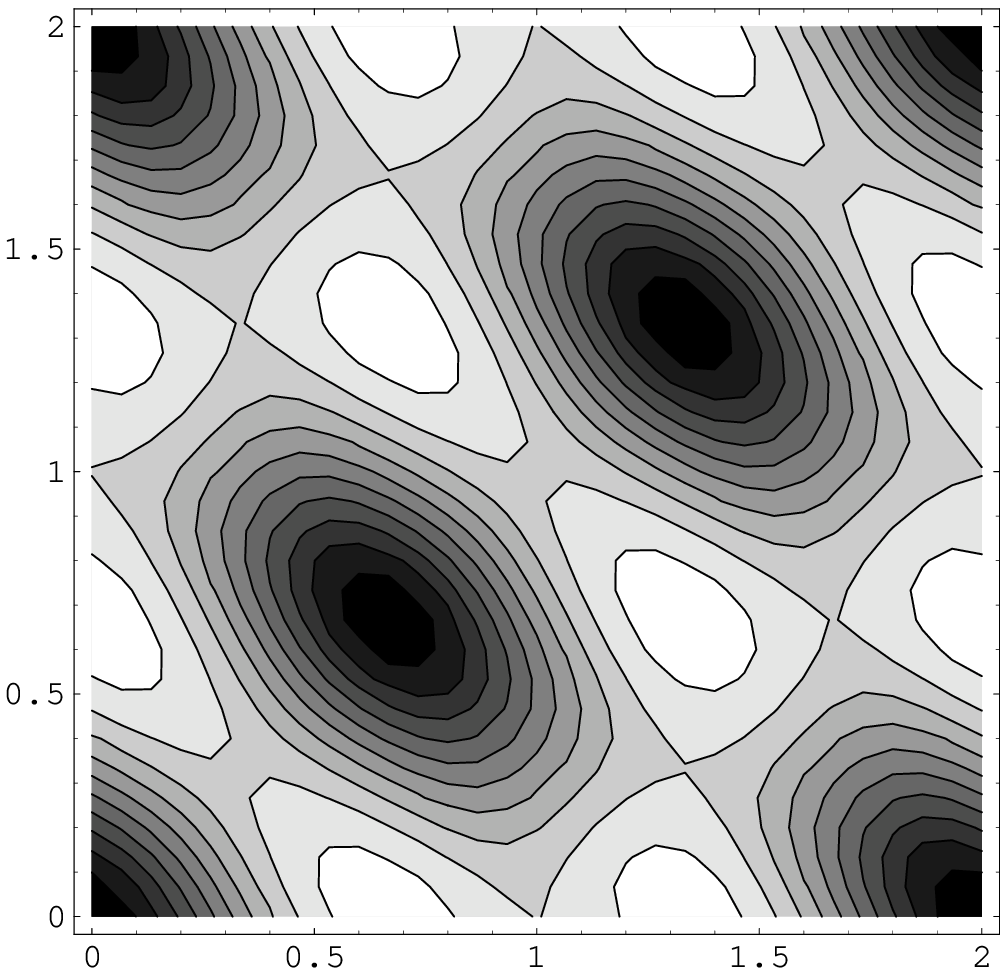}}
\epsfxsize=5cm
\centerline{(b)\epsfbox{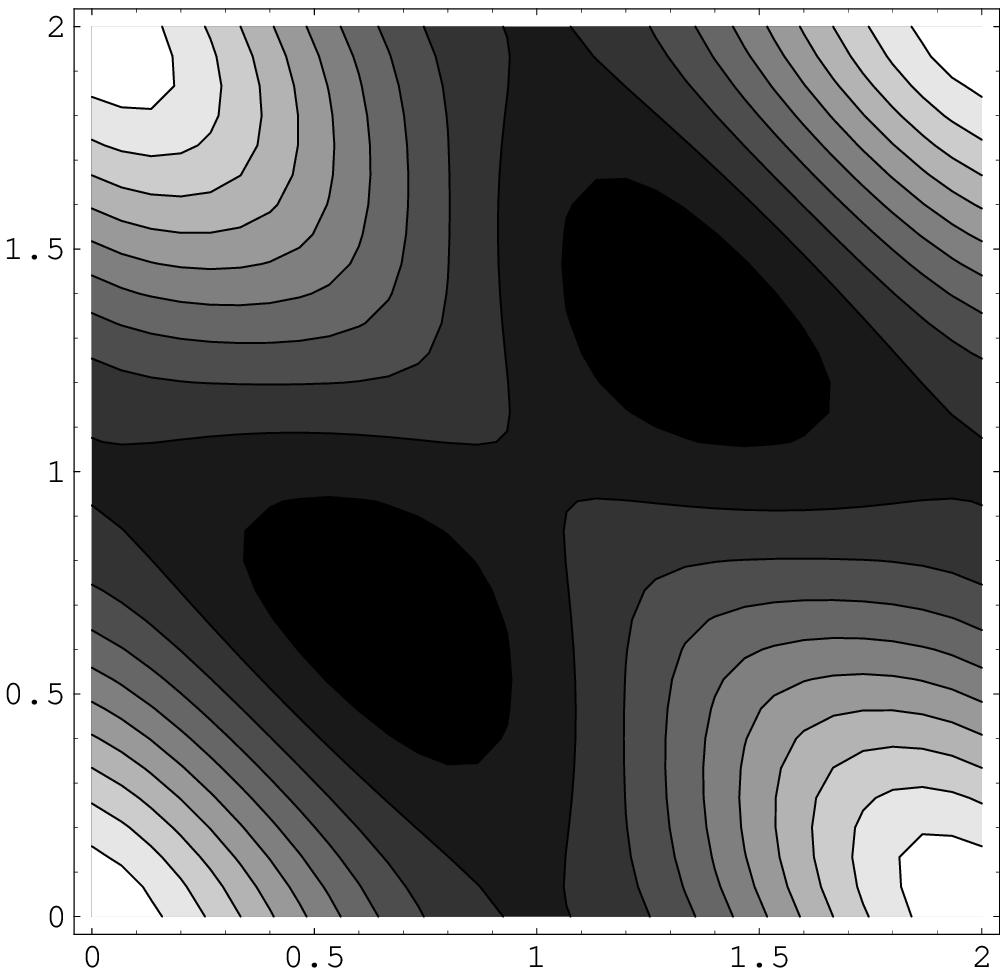}}
\epsfxsize=5cm
\centerline{(c)\epsfbox{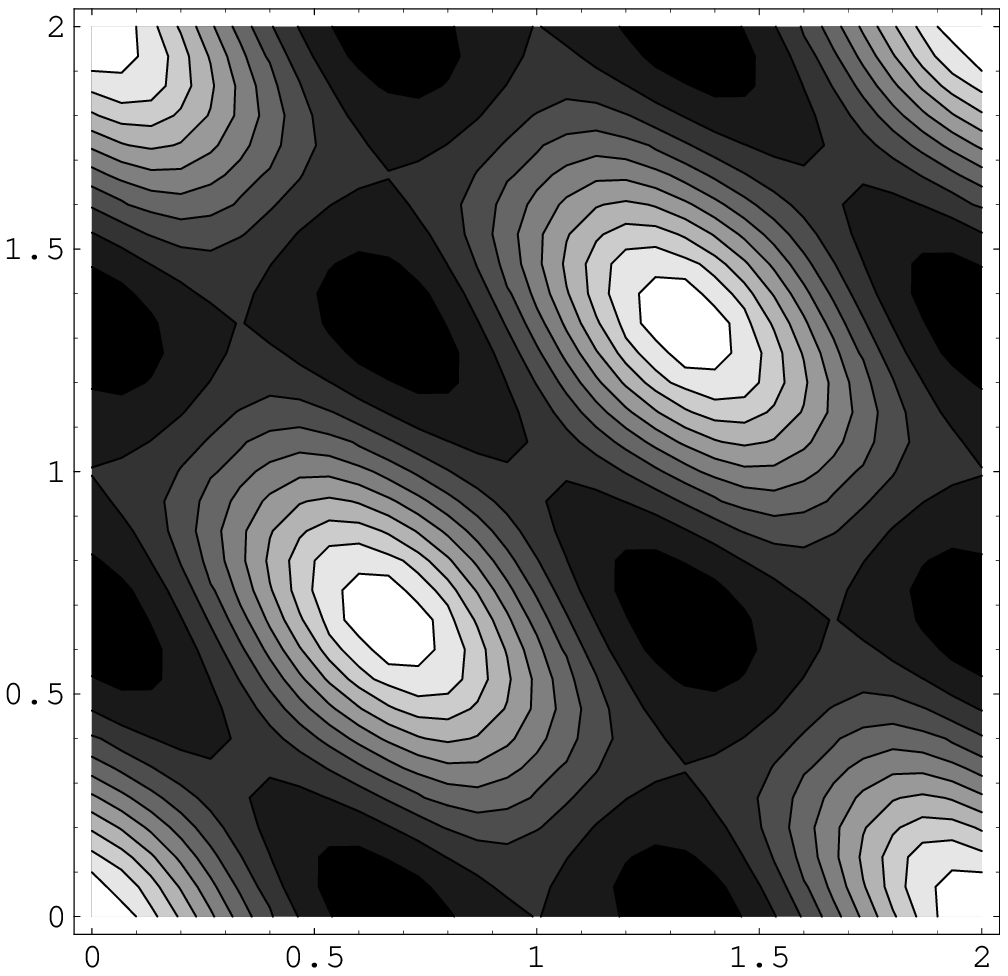}}
\caption{
Plots of the $\Veff$.
(a) $\Veff^{\mathrm{g+gh}}(\theta_1,\theta_2)$ (for $\Nad=\Nf=0$),
(b) $\Veff^{\mathrm{f}}   (\theta_1,\theta_2)$ (for $\Nad=1/2$, $\Nad=1$), and
(c) $\Veff^{\mathrm{ad}}  (\theta_1,\theta_2)$ (for $\Nad=1$, $\Nf=0$).}
\end{figure}
We can see that $\Veff^{\mathrm{g+gh}} + \Veff^{\mathrm{f}}$ does not
have a minima at $\theta_1=\theta_2=0$ for sufficiently large $\Nf$, and
$\Veff$ has global minima at points such as $(\theta_1,\theta_2) =
(2\pi/3,2\pi/3)$, as pointed out by Hosotani.\cite{Hosotani83} The
points $(\theta_1,\theta_2)=(0,0)$ and $(2\pi/3,2\pi/3)$ are all $SU(3)$
symmetric. On the other hand, $\Veff^{\mathrm{ad}}$ has minima at other
points, for example at $(\theta_1,\theta_2)=(0,2\pi/3)$, where $SU(3)$
symmetry is broken.

Our main interest here is whether the breaking
$SU(3)\rightarrow SU(2) \times U(1)$ is possible. Such a study will be
helpful when we apply the method to more realistic GUT theories, where
partial breaking like $SU(5)\rightarrow SU(3) \times SU(2) \times U(1)$
is favored. When $\Nad=1$ and $\Nf=3$, $\Veff$ seems to have a minimum at
$(\theta_1,\theta_2,-(\theta_1+\theta_2))=(0,\pi,-\pi)$ and its
permutations (Fig. 3). 
\begin{figure}[htbp]
\epsfxsize=5cm
\centerline{
%\resizebox{6cm}{!}{\includegraphics{su3nf3.eps}}
\epsfbox{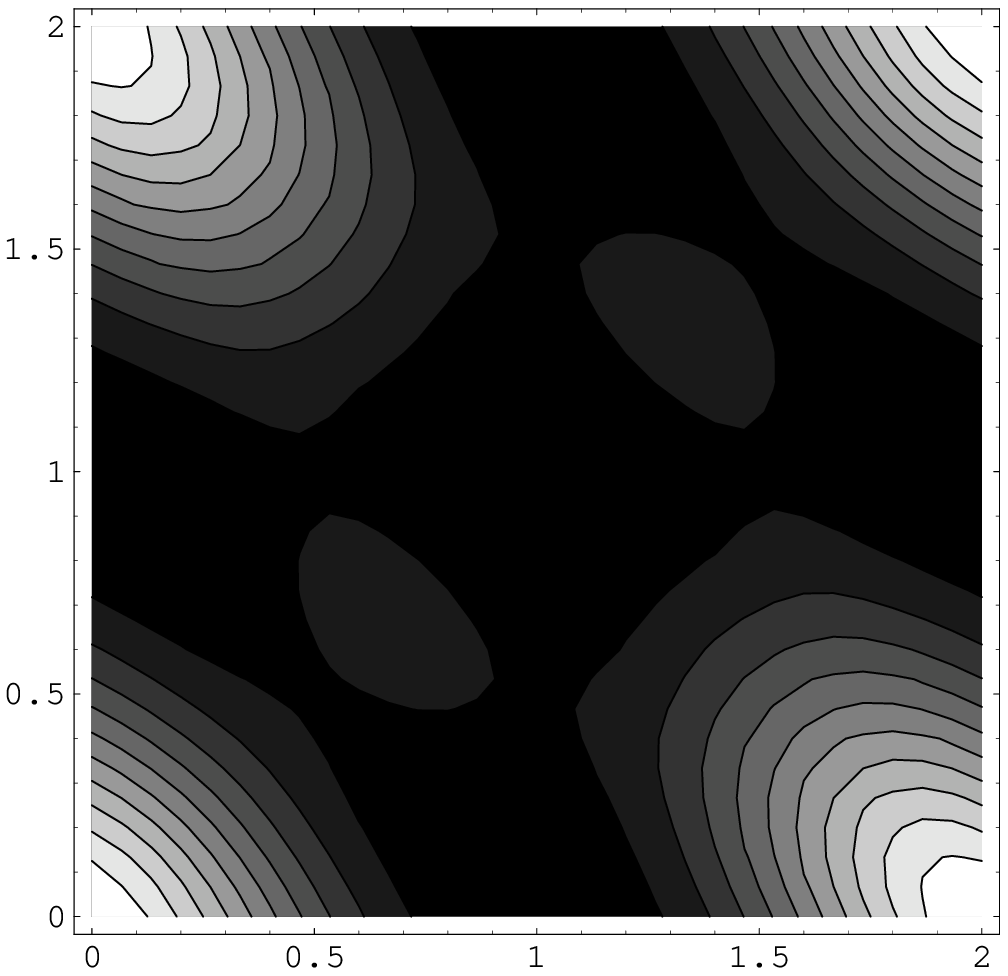}
}
\caption{$\Veff$ plot ($\Nad=1$, $\Nf=3$).}
\end{figure}

To determine the position of the global minimum, evaluating the Hessian
of $\Veff$,
\begin{equation}
H_{\Veff}(\theta_1,\theta_2;\Nad,\Nf)_{i,j}
\equiv
\left.
\frac{\partial^2 \Veff(\theta_1,\theta_2;\Nad,\Nf)}{\partial \theta_i \partial \theta_j}
\right|_{\theta_1,\theta_2},
\end{equation}
and its eigenvalues and eigenvectors is useful.  By utilizing such a 
method, we study how the gauge symmetry breaks by changing $\Nf$ with
$\Nad = 1$ for brevity.
\begin{enumerate}

\item[(1)]{ $ 0 \le \Nf < 3 $ }

For $\Nf=0$, the minima of the potential coincides with those of
$\Veff^{\mathrm{ad}}$.
As $\Nf $ increases, such minima move to the points where $SU(2) \times U(1)$
symmetry is realized.

\item[(2)]{ {$ 3 \le \Nf < 9 $} }

When $\Nf=3$, global minima are at the points $(0,\pi,-\pi)$ (and its
permutations). It also should be noted that the Hessian of $\Veff$
exactly vanishes at such points:
\begin{eqnarray}
\det H_{\Veff}(0,\pi;\Nad=1,\Nf=3)
\propto
- (6\Nad - 3 -\Nf )^2 |_{\Nad=1,\Nf=3}
= 0.
\end{eqnarray}
          This means that the mass of the adjoint Higgs as a curvature
          of the effective potential vanishes exactly. Then as $\Nf$
	  increases, global minima of $\Veff$ leave from $SU(2) \times
	  U(1)$ symmetric points ($(0,0)$, $(0,\pi)$, $(\pi,0)$) and
	  move to the $SU(3)$ symmetric points $( 2\pi/3, 2\pi/3 )$,
	  keeping $SU(2)\times U(1)$ symmetry.

\item[(3)]{$ 9 \le \Nf$} 

When $\Nf \ge 9$, global minima are at $SU(3)$ symmetric (non-trivial)
points, such as $( 2\pi/3, 2\pi/3 )$.

\item[(4)]{Negative $\Nf$}

As $\Nf$ decreases, the global minima move to the trivial VEV, such as
$(0,0)$. When $\Nf \le -3$, $SU(3)$ symmetry is recovered.
\end{enumerate}

\subsubsection{Summary}
The results obtained above can be summarized as follows:
\begin{equation}
SU(3)
\rightarrow 
\left\{
\begin{array}{ccccccc}
SU(3)            &\mbox{for}  &            &    & \Nf &\leq& 3 - 6\Nad, \\
U(1)^2           &\mbox{for}  & 3-6\Nad    &<   & \Nf &<   & 6\Nad - 3, \\
SU(2)\times U(1) &\mbox{for}  & 6 \Nad-3   &\le & \Nf &<   & 18\Nad-9, \\
SU(3)            &\mbox{for}  & 18 \Nad-9  &\le & \Nf.&    &
\end{array}
\right. 
\end{equation}
Hence all possible symmetry breaking patterns turn out to be realized.

It is interesting to note that in the case $ \Nf = 6\Nad - 3$, not only
does the Wilson line allow $SU(2)\times U(1)$ symmetry, but also all
components of the Hessian at $SU(2)\times U(1)$ symmetric points vanish. 
This means that by the Hosotani mechanism, $SU(3)$ symmetry breaks into
$SU(2) \times U(1)$, and at this minimum, the effective potential has a
vanishing second derivative, which again suggests that the bosons remain
massless.

We must remark that we get such a massless state without any
fine-tuning of the parameters because the equality $\Nf - 6\Nad + 3 = 0$
is exactly satisfied without any fine-tuning of parameters, as both
$\Nad$ and $\Nf$ are discrete numbers.

\subsubsection{$d=5$, $d=6$ cases}

As pointed out for the $SU(2)$ model, the situation of symmetry breaking
depends also on the dimensionality of space-time.

Dependences on the matter content is more complicated for $d=5$ and
$d=6$ case.  For example, in the $d=5$ case, $\det H_{\Veff} (2\pi/3,
2\pi/3) > 0$ suggests that $(2\pi/3,2\pi/3)$ is at least a local minimum
of $\Veff$. However, for $\Nf = (108\Nad-81)/8$, as shown in Fig. 4,
such a point is not the global minimum.
% figure 4
\begin{figure}[htbp]
\epsfxsize=5cm
\centerline{
%\resizebox{6cm}{!}{\includegraphics{su3dg.eps}}
\epsfbox{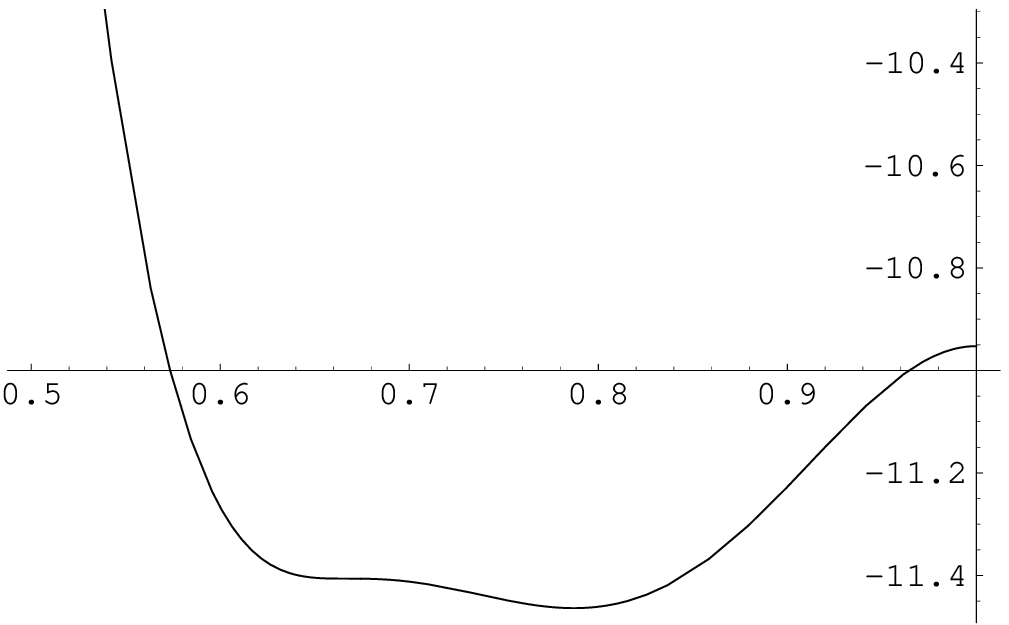}
}
\caption{Plot of $\Veff(\theta,\theta)$ for $\theta=\pi/2$ to $\pi$
 with $\Nad=1$ and $\Nf=27/8$.}
\end{figure} 
$SU(3)$ symmetry is recovered for sufficiently large $\Nf$, though the
critical value at which $SU(3)$ symmetry is recovered is more ambiguous
than in the $d=4$ case. We see that $SU(3)$ is recovered at least when
$\Nf > 4$ for $\Nad=1$ case.

For $d=6$, we have
\begin{equation}
\mbox{SU(3)}
\rightarrow 
\left\{
\begin{array}{ccccccc}
\mbox{SU(3)}  &\mbox{for} &            & & \Nf &\leq& 3 - 6 \Nad, \\
\mbox{U(1)}^2 &\mbox{for} & 3 - 6\Nad  &<& \Nf &<   & (18\Nad - 9)/7.
\end{array}
\right.
\end{equation}
The critical value at which $SU(2)\times U(1)$ is enlarged to $SU(3)$ is
ambiguous, as in the case of $d=5$.  The transition $U(1)^2 \rightarrow
 SU(2)\times U(1)$ occurs at $(18\Nad - 9)/7 = \Nf $, and $SU(3)$
symmetry is recovered at least at $\Nf=8$.

It also must be pointed out that in the $d=5$ and $6$ cases, the
curvatures at $SU(2)\times U(1)$ symmetric points never vanish, unlike
in the $d=4$ case.

\subsection{$F_d(\theta)\rightarrow \cos\theta$ (F2C) approximation}

We found that gauge symmetry breaking depends not only on the matter
content of the theory but also on the dimensionalities of the
space-time.  This is partially because the factors $2^{[2/d]}$ and
$-(d-2)$ depend on the dimensionality $d$. The other reason is that the
form of the function $F_d(\theta)$ depends on $d$. Generally,
$F_d(\theta)$ becomes close to $\cos\theta$ as $d$ becomes large. Thus
we can approximate $F_d(\theta)$ by $\cos\theta$ for sufficiently large
$d$.

By taking $F_d(\theta) \rightarrow \cos\theta$ (F2C, for short) as an
approximation, we can more easily calculate and evaluate the $\Veff$ for
the case of $SU(3)$ with $d=10$.

To summarize the results,
\begin{equation}
SU(3)
\rightarrow 
\left\{
\begin{array}{ccccccc}
SU(3)            & \mbox{for}     
&                &   & \Nf &\leq& (3 - 12 \Nad)/2 ,\\
U(1)^2           & \mbox{for}
& (3 - 12\Nad)/2 &<  & \Nf &<   & (4\Nad - 1)/2 ,\\
SU(2)\times U(1) & \mbox{for}
& (4\Nad -1)/2   &\le& \Nf &<   & 16\Nad - 4 ,\\
SU(3)            & \mbox{for}
& 16\Nad -4      &\le& \Nf, &    &                     
\end{array}
\right.
\end{equation}
where the critical value for the $SU(2) \times U(1) \rightarrow SU(3)$
transition is obtained by simply comparing the values of
$\Veff(2\pi/3,2\pi/3)$ and $\Veff(\pi,\pi)$, because minima of $\Veff$
cannot be realized except at these points for $(4\Nad-1)/2 < \Nf <
16\Nad-4$ in this approximation.

%%%%%%%%%%%%%%%%%%%%%%% section 3 %%%%%%%%%%%%%%%%%%%%%%%%%%%
\section{Discussion}

In this paper, it has been shown that in the Hosotani mechanism, the
breaking pattern of gauge symmetry depends on the choice of the
representation of matter fields.  The manner in which the matter in the
fundamental and adjoint representations contribute to the gauge symmetry
breaking has been clearly shown in the $SU(2)$ case.

By explicit calculation we have also shown for the $SU(3)$ model that we
can realize not only breaking of $SU(3)$ gauge symmetry but also
breaking into $SU(2)\times U(1)$, i.e., partial gauge symmetry breaking,
by suitably selecting the combination of $\Nf$ and $\Nad$.  This feature
may be very desirable in constructing more realistic GUT models where a
breaking into a subsymmetry with the same rank, such as $SU(5)\rightarrow
SU(3)\times SU(2)\times U(1)$, is needed. Ho and Hosotani's
model\cite{hohosotani} contains only an adjoint matter field, and the
$SU(N)$ gauge group breaks only into $U(1)^{N-1}$ with periodic
boundary conditions, and nontrivial boundary conditions were required for
$SU(5)$ to break into $SU(3)\times SU(2)\times U(1)$.

Including the desirable feature mentioned just above, our model has the
following advantages, when considered as a prototype model of GUT.

\begin{enumerate}
\item Our model contains fermions with the fundamental representation in
      addition to the adjoint representation, while the models discussed
      by Hosotani et al. have only fermions with the adjoint
      representation. The situation we discussed may be understood as
      more realistic, since matter with the fundamental representation
      is needed in GUT.
\item It should be stressed that in our model, once the representation
      of the matter field is fixed, there is no arbitrariness in $\Veff$
      and in the breaking pattern of gauge symmetry, whereas in ordinary
      GUT they depend crucially on the values of many parameters of
      self-coupling of the Higgs, such as the {\bf 24} representation of
      the Higgs in $SU(5)$. We have shown that in our model the VEVs of
      Higgs (as extra space components of the gauge boson) are
      dynamically (and uniquely) chosen as the bottom of $\Veff$. As a
      bonus we have also shown that our 4-dimensional gauge theory must
      live at a ``flat bottom'' of $\Veff$, namely that the Higgs does
      not suffer from the large finite mass correction encountered in
      our previous work, in which an Abelian gauge theory is
      discussed. \cite{HIL98}
\end{enumerate}

To construct a realistic model, we must investigate models with larger
gauge groups, such as $SU(5)$, and with other representations, such as
{\bf 10} of $SU(5)$. Moreover, the compactified manifold may be a larger
and more complicated one, such as $T^n$($n \ge 2$). The analysis may
become somewhat more complicated. The F2C approximation mentioned in
this paper should help to simplify the analysis.

\subsection*{Acknowledgements}

I would like to thank Professor C.~S.~Lim for useful comments,
discussions and careful reading of this manuscript.  I am also grateful
to Professor T.~Inami, Professor K.~Shiraisi, Professor M.~Sakamoto, Dr. 
K.~Takenaga and Dr. M.~Tachibana for their valuable suggestions on this
reserch.

%%%%%%%%%%%%%%%%%%%%%%%%%%%%%%%%%%%%%%%%%%%%%%%%%%%%%%%%%%%%%%%%%%%%%%%%
%                              References
% %%%%%%%%%%%%%%%%%%%%%%%%%%%%%%%%%%%%%%%%%%%%%%%%%%%%%%%%%%%%%%%%%%%%%%
\setlength{\baselineskip}{24pt}

\end{document}